\documentclass{iaus}
\usepackage{graphicx}

\begin {document}
\title[B/P bulges, buckling and bars]{Boxy/Peanut bulges, vertical buckling and galactic bars} 

\author[Martinez-Valpuesta \& Athanassoula]   
{Inma Martinez-Valpuesta$^{1,2}$ and E. Athanassoula$^2$}

\affiliation{$^1$Instituto de Astrof\'isica de Canarias, C/Via L\'actea, s/n, 38200, La Laguna, Tenerife, Spain \\[\affilskip]
$^2$LAM, OAMP, 2 Place Le Verrier, F-13004 Marseille, France}

\pubyear{2007}
\volume{245}  
\pagerange{--}
\date{?? and in revised form ??}
\setcounter{page}{1}
\jname{Proceedings Title IAU Symposium}
\editors{A.C. Editor, B.D. Editor \& C.E. Editor, eds.}
\maketitle

\begin{abstract}
Boxy/peanut bulges in disk galaxies have been associated to stellar bars. In
this talk, we discuss the different properties of such bulges and
their relation with the corresponding bar, using a very large sample of a few 
hundred numerical N-body simulations.
We present and inter-compare various methods of measuring the
boxy/peanut bulge properties, namely its
strength, shape and possible asymmetry. Some
of these methods can be applied to both simulations and observations. 
Our final goal is to get correlations that will allow us to obtain
information on the boxy/peanut bulge for a galaxy viewed face-on as
well as information on the bars of galaxies viewed edge-on.
\end{abstract}

\firstsection

\section{Introduction}

Simulations have 
shown that bars are not vertically thin morphological
features, but have a considerable vertical extent and a 
vertical structure, known as the Boxy/Peanut bulges (hereafter B/P;
Combes \& Sanders 1981, Combes et al. 1990). Comparisons between
observations and $N$-body simulations have established this
direct connection firmer (Athanassoula 2005 and references
therein). Furthermore, observations have shown that both bars and B/P
bulges are 
quite predominant in disc galaxies and that the corresponding
frequencies are in good agreement with the link between the two
structures (L\"utticke, Dettmar \& Pohlen 2000). 

We measure the peanut properties in a large sample of several hundred
$N$-body simulations ran by one of us (EA) for different
purposes. More information on these simulations and on their
properties can be found in Athanassoula \& Misiriotis (2002) and
Athanassoula (2003, 2007). In particular, we seek correlations between
the properties of the bar and the properties of the B/P bulge. 

\section{Methods for measuring bar and peanut strength and correlations}
In order to measure the bar strength, we use standard Fourier
decomposition and take the amplitude of the second ($m$ = 2)
component. 

\begin{equation}
C_{m,r} (R) = |\sum_{j=1}^{N_s}~m_j e^{i m \theta_j}|, ~~~~m=2,
\label{eq:Cmr}
\end{equation}

\noindent
where $m_j$ and $\theta_j$ are the mass and azimuthal angle of
particle $j$. This can be done either globally or as a function of radius. In the
former case, the summation is carried out over all particles in the
disc, while in the latter $N_s$ is the number of particles in a
given cylindrical shell of radius $R$ and $C_{m,r} = C_{m,r} (R)$. 

The vertical asymmetry and the strength of the B/P can be measured in
a similar way, by  

\begin{equation}
C_{m,z} = | \sum_{j=1}^{N_s}~z_j e^{i m z_j / (5 z_0)}|, ~~~~~~~m=1, 2,
\label{eq:Cmz}
\end{equation}

\noindent
where $N_s$ can be either the number of particles in the disc
component, or the number of particles per vertical column or cut (in
which case $C_{m,z} = C_{m,z} (R)$) and $z_0$ is the scale height of the
initial exponential disc. In Fig.~1 we apply these two definitions to
two simulation snapshots, one with a boxy bulge and the other with a
peanut, or X-shaped bulge.  

\begin{figure}
\begin{center}
\includegraphics[scale=0.45]{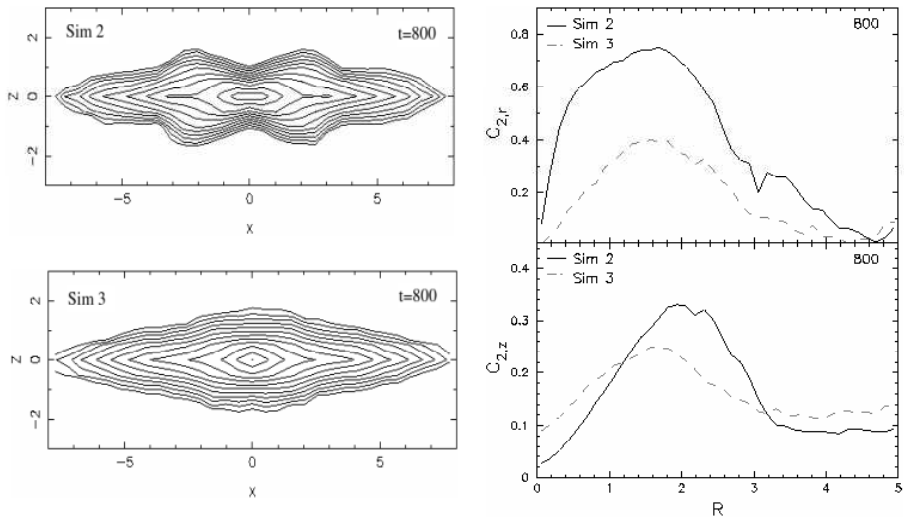}
\caption{{\it Right panel:} Edge-on view of two different simulations. 
{\it Left panel:} Strength of the bar (top panel) and 
of the B/P bulge (lower panel) vs. radius, for two snapshots, 
one corresponding to a boxy bulge (dashed line) and the other 
to a peanut (full line).
}
\end{center}
\end{figure}

We have also applied statistical methods, based on the distribution of
the $z$ coordinates of the particles in a given vertical cut
(perpendicular to the equatorial 
plane) of a snapshot seen edge-on, with the bar viewed side-on. We use
the median of the absolute values and the standard deviation ($\sigma_z$) 
to measure the strength of the peanut, the kurtosis to measure its shape
and the skewness and the mean to measure the strength of the buckling
event, i.e. the asymmetry during the vertical instability.   
 
We inter-compared the different methods of measuring the B/P strength
and found very strong correlations, allowing us to use these methods
indiscriminately. This is important, particularly in view of their
application to real, highly inclined galaxies. 

\section{Detecting the buckling events in simulations}

We use three different methods to determine the strength of the
buckling event and the time at which it occurs. We again use the Fourier
decomposition, but now the snapshot is viewed edge-on and we use the
amplitude of the first coefficient in the Fourier decomposition. This
first coefficient gives a measure of the asymmetry: 

\begin{equation}
A_{1,z} = \frac {1}{M_k \pi} |\sum_j m_j e^{im \phi_j}|, 
\label{eq:A1z}
\vspace{-0.1cm}
\end{equation}

\noindent
where $\phi_j$ is the angle of particle $j$ measured in the ($x,z$)
plane, with the major axis of the bar, positioned in principle, in the $x$-axis. To make this coefficient
more sensitive to a buckling, we 
usually exclude from the summation disk particles that are not in the bar, as
discussed in Martinez-Valpuesta et al. (2006). The left panels of Fig.~2
show the time evolution of  $A_{1,z}$ for four different
simulations. The over plotted lines correspond to different position
angles of the bar (from $0^\circ$ to $90^\circ$) and
to different inclinations of the galaxy (from $i=80^\circ$ to
$i=90^\circ$). Notice that the peaks in $A_{1,z}$ are visible in all
these orientations. We would like to point out the panel showing Sim~4, where two peaks are seen. These peaks correspond to the two buckling events that many of our simulations undergo (also seen in Martinez-Valpuesta et al. 2006).
As mentioned above, we can also measure the strength of the buckling
using statistics on the distribution of the $z$
coordinates of the simulation particles, i.e. measure
the mean ($<z>$) and the skewness ($S_z$). The value of these
statistical moments vs. radius for a given snapshot, i.e. simulation
and time, is shown in Fig.~2, right bottom panel. The absolute
values of these parameters will be maximal at the time when the
buckling event is happening. Fig.~3 compares the results of our three
methods for measuring the buckling strength and shows that they
correlate with each other.

\begin{figure}
\begin{center}
\includegraphics[scale=0.4]{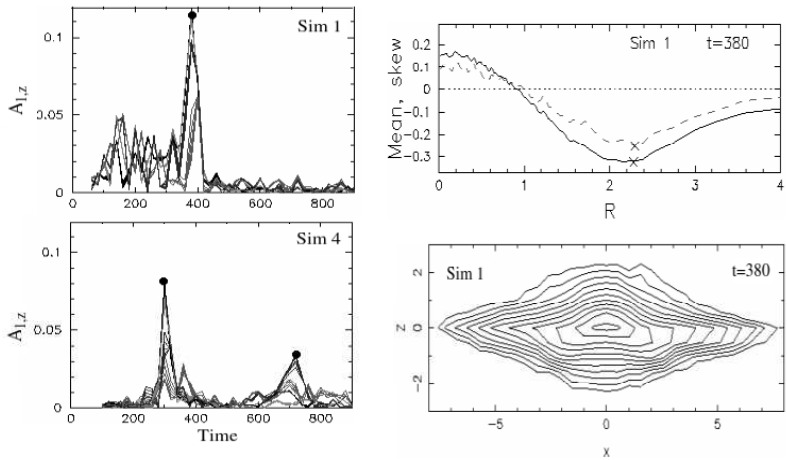}
\caption{{\it Left panel}: $A_{1,z}$ for two different simulations. The different lines in each plot correspond to the different orientations in position angle of the bar, and inclination of the galaxy.
{\it Right panels}: In the upper panel, we show the the mean and skewness for the same simulation at the time of the buckling. In the lower panel we show the edge-on view at the time of the buckling for Sim~1.
}
\end{center}
\end{figure}

\begin{figure}
\begin{center}
\includegraphics[scale=0.7]{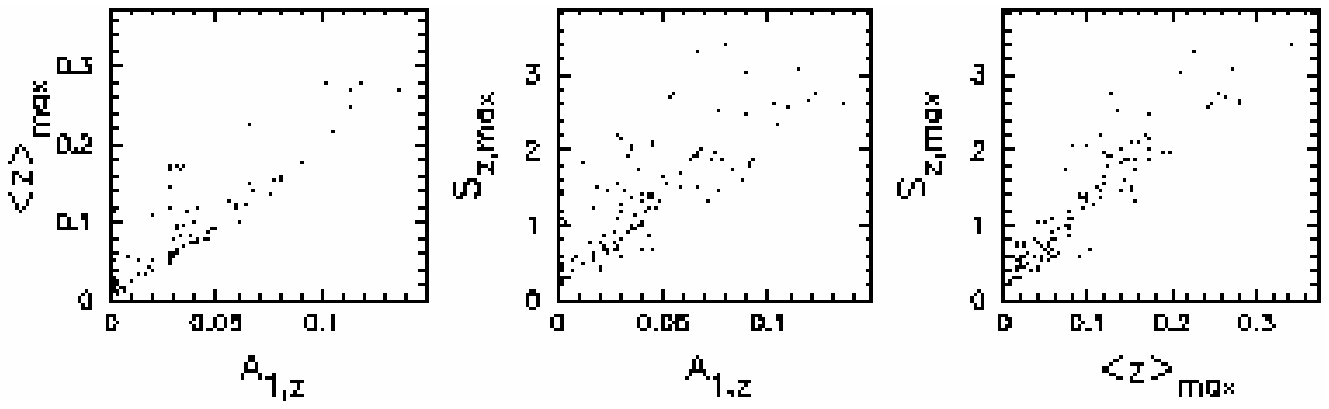}
\caption{Correlations between the results of the three different methods we
  use to calculate the strength of the buckling event.
}
\end{center}
\end{figure}

More important, we find correlations
between both the strength and the shape of the B/P bulge with the
strength of the bar (Fig.~4). The shape of the B/P bulge is given by
the minimum kurtosis.  
In this figure we show that strong bars have buckled more times, and
achieved a stronger B/P bulge.  

\section{Conclusions}

We presented several methods to calculate the strength of the bar and
the strength, shape and asymmetry of the B/P bulge and found strong
correlations between their results. The most important correlation
relates the strength of the bar with the strength of the B/P bulge,
the strongest bars having the strongest peanuts. We also find that the
strength of the peanut depends on the number of buckling episodes it
underwent, the strongest bars having undergone more buckling episodes
(Fig. 4). Finally, we find a very interesting result about $C_{2,z}(R)$,
i.e. about the shape of the radial density profiles along cuts
perpendicular to the equatorial plane. For strong bars, having a
strong peanut or X-shaped bulge, this profile is more flat-topped,
while for weaker bars, with more boxy-like bulges, it is more
peaked. All the results summarised here are
discussed in length by Athanassoula \& Martinez-Valpuesta (2007, in
preparation). 

\begin{figure}
\begin{center}
\includegraphics[scale=0.35]{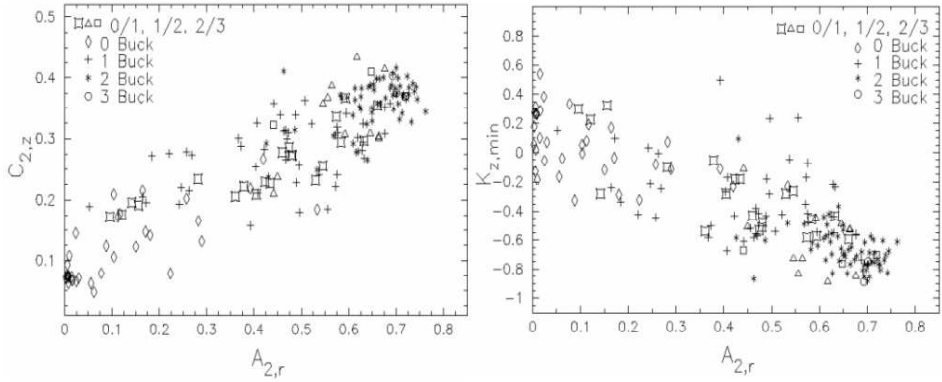}
\caption{Correlations between bar and peanut properties. 
Each symbol corresponds to one simulation. The type of symbol is
related to the number of buckling events suffered by the bar during
its evolution. {\it Left panel}: Strength of the B/P bulge measured
with our Fourier based method vs. the strength of the bar. {\it Right
  panel}: Shape of the B/P bulge (i.e. shape of the radial density
profiles along cuts perpendicular to the equatorial plane, measured by
the minimum of the kurtosis) plotted as a function of bar strength. 
}
\end{center}
\end{figure}
\acknowledgements This work has been partially supported by the Peter and Patricia Gruber Foundation Fellowship.

\end{document}